\documentclass[12pt]{article}
\usepackage{amssymb,latexsym,epsf} 
\usepackage{epsf}  
\usepackage{a4wide}

\def\be{\begin{equation}}
\def\ee{\end{equation}}
\def\bea{\begin{eqnarray}}
\def\eea{\end{eqnarray}} 
\def\nn{\nonumber \\}

\def\part{\partial}
\def\tfrac#1#2{{\textstyle{#1\over #2}}}
\def\half{\tfrac{1}{2}}

\def\hg{{\hat g}}

\def\tg{{\tilde g}}

\def\makeatletter{\catcode`\@=11}
\makeatletter
\def\mathbox#1{\hbox{$\m@th#1$}}%
\def\math@ccstyles#1#2#3#4#5#6#7{{\leavevmode
      \setbox0\mathbox{#6#7}%
      \setbox2\mathbox{#4#5}%
      \dimen@ #3%
      \baselineskip\z@\lineskiplimit#1\lineskip\z@
      \vbox{\ialign{##\crcr
             \hfil \kern #2\box2 \hfil\crcr
             \noalign{\kern\dimen@}%
             \hfil\box0\hfil\crcr}}}}
\def\mathaccstyles{\math@ccstyles\maxdimen}
\def\maththroughstyles{\math@ccstyles{-\maxdimen}}
\def\unitmatrixDT%
 {\maththroughstyles{.45\ht0}\z@\displaystyle {\mathchar"006C}\displaystyle 1}
\begin{document}

\rightline{IFT-UAM/CSIC-00-11}
\rightline{hep-th/0003002}
\rightline{\today}
\vspace{1truecm}


\centerline {\Large \bf Brane World with Bulk Horizons}
\vspace{1truecm}

\centerline{
    {\bf C\'esar G\'omez,}\footnote{E-mail address: 
                                  {\tt cesar.gomez@uam.es}}
    {\bf Bert Janssen}\footnote{E-mail address: 
                                  {\tt bert.janssen@uam.es}}
    {\bf and} 
    {\bf Pedro J. Silva}\footnote{E-mail address: 
                                  {\tt psilva@delta.ft.uam.es}}}
\vspace{.4truecm}
\centerline{{\it Instituto de F{\'\i}sica Te{\'o}rica, C-XVI,}}
\centerline{{\it Departamento de F{\'\i}sica Te{\'o}rica, C-XI, }}
\centerline{{\it Universidad Aut{\'o}noma de Madrid}}
\centerline{{\it E-28006 Madrid, Spain}}
\vspace{2truecm}

\centerline{\bf ABSTRACT}
\vspace{.5truecm}

\noindent
A brane world in the presence of a bulk black hole is constructed. The
brane tension is fine tuned in terms of the black hole mass and
cosmological constant. Gravitational perturbations localized
on the brane world are discussed.

\newpage

\noindent
{\bf 1. Introduction}
\vspace{.3cm}

A brane world with induced four dimensional gravity was first introduced in 
\cite{RS1,RS2} on the basis of a $AdS_{5}$ bulk geometry. In this
scheme normalizable gravitational zero modes are allowed due to the ultraviolet
cutoff induced by the brane wall. The dilaton field is constant and
the holographic degrees of freedom on the wall define a conformal field
theory coupled to gravity \cite{W1}. In a series of recent papers 
\cite{Youm,CS1,AD,GJS,CS2} this framework was extended to the non conformal 
case i.e to dilatonic domain walls. In these cases, both with vanishing and 
non vanishing cosmological constant, we observe the phenomena of induced four
 dimensional gravity, however a naked curvature singularity is induced by
the non constant dilaton in the bulk at a finite proper distance from
the brane wall. The physics interpretation of such a singularity
from the four dimensional point of view is still an open problem.

In this letter we will look for a Randall-Sundrum scenario but this time
in the bulk geometry of a real black hole with the singularity inside
a trapped surface. A similar analysis was with a non static Ansatz was first 
by \cite{Kr, KK} in the framework of cosmological models. We will work out the 
static case in a Schwarzschild-AdS bulk metric. This will correspond to a brane
world in a thermal bath at the Hawking temperature.
The first question we will address would be the fine tuning relations
between the brane wall tension and the parameters $\Lambda$ and $M$ 
characterizing the Schwarzschild-AdS metric.  These fine tuning relations would
be obtained by solving the corresponding jump equations once we introduce
the wall as an ultraviolet cutoff analogously to the AdS case. Since
our space is asymptotically AdS this cutoff could be enough to induce
four dimensional gravity on the wall in terms of normalizable
graviton zero modes. 
\vspace{.5cm}

\noindent
{\bf 2. Construction of the solution}
\vspace{.3cm}

Our starting point is the following five-dimensional action of gravity 
in the presence of a cosmological constant $\Lambda$ with a domain wall
 source term given by:
\be
S=\frac{1}{\kappa} \int d^4x\ dy \ \sqrt{|g|} \ \Bigl[{\cal R} -  \Lambda \Bigr] \ 
             + \int d^4x \sqrt{|\tg|} \ V_0,
\label{lagran}
\ee 
where $V_0$ is the tension of the brane and 
$\tg_{mn}=g_{\mu\nu}\delta^\mu_m\delta^\nu_n$ the induced metric on the brane.

We are interested in a solution resembling the Schwarzschild-AdS solution, 
which is given by \cite{HP}:
\be
ds^2= (1+R^{-2}r^2-\tfrac{2M}{r^2})\ dt^2 
   - \frac{1}{(1+R^{-2}r^2-\tfrac{2M}{r^2})}dr^2 - r^2 d\Omega_3^2 \
.
\ee
where $R=12 \Lambda^{-1}$ and $M$ is basically the black hole mass. In
order to apply the Randall-Sundrum program to this type of metrics, we
 consider a new set of coordinates defined by, 
\be
dz = \frac{1}{\sqrt{1+R^{-2}r^2-\tfrac{2M}{r^2}}}\  dr \ ,
\label{coordtransf}
\ee
therefore ending up in a holographic-like frame
\be
ds^2= A^2(z) dt^2 + B^2(z) d\Omega^2_3 - dz^2 \ ,
\label{Ansatz}
\ee
where $A(z)$ and $B(z)$ are function of the holografic coordinate $z$, 
implicitly given by
\be
A(r)= \sqrt{1+R^{-2}r^2-\tfrac{2M}{r^2}} \ , 
\hspace{2cm}
B(r)=r\ ,
\label{functions}
\ee
with $r$ a function of $z$, defined by the relation (\ref{coordtransf}).
Using this Ansatz (\ref{Ansatz}) in the corresponding equations of motion of 
(\ref{lagran}), we get the following system of equations:
\bea
&& B^{-2} \Lambda \ - \  B^{-2}(B')^2 \ - \ A^{-1}A'B^{-1}B'=0 \ , \nn
&& - B^{-2} + B^{-1}B'' + B^{-2}(B')^2 +\Lambda 
               + \tfrac{1}{6}\kappa V_0 \delta(z)=0 \ , \\
&& -1 + \half B^2 \Lambda + A^{-1}B^2A'' +2BB'' + (B')^2 + 2A^{-1}A'BB'
            + \tfrac{1}{2}\kappa V_0 \delta(z)=0 \ , \nonumber
\label{eom}
\eea
where `` $'$ '' means  derivatives with respect to $z$ and have used the fact
 that $d\Omega^2_3$ is a three-dimensional sphere of radius one. 

To obtain the desired solution, we use the fact that (\ref{functions})
is a solution of the action without source term, therefore in our case 
the solution to the full equations (\ref{eom}) is given by the functions
(\ref{functions}) with a modification of the relation between $z$ and $r$, 
defined as follows
\be
|\bar{z}| = z_0 - \tfrac{R}{2} \ \log \ \Bigl|
                  \frac{2R^{-1}rA+ 2R^{-2}r^2 +1}{\sqrt{1+ 8MR^{-2}}} \Bigr|    
\label{bla}
\ee
where $\bar{z}=z_0-z$, which translates into:
\be
d|\bar{z}| = - \frac{1}{\sqrt{1+R^{-2}r^2-\tfrac{2M}{r^2}}}\  dr \ .
\label{coordtransf2}
\ee
The modulus of the radial coordinate $\bar{z}$ runs between the the
position of the brane at $\bar{z}=0$ and the black hole horizons at $|\bar{z}|=z_0$. In this coordinates this means that the brane is located
somewhere between the horizon and the boundary. 

By solving the jump equations we find that the brane tension $V_0$ is given 
in terms of the cosmological constant $\Lambda$, the black hole mass $M$
and $r(\bar{z}=0)$, provided the brane is located at the origin in $\bar{z}$ coordinates, 
by the following relation:
\be
V_0= \frac{6}{\kappa r(0)}\sqrt{1+R^{-2}r(0)^2-\tfrac{2M}{r(0)^2}}
\ee
If the black hole horizon is smaller than the AdS radius $M<R^2$, 
we could choose to do the cutoff at $r(\bar{z}=0)= R$ and the above formula 
reduces to:
\be
V_0= \frac{6}{\kappa R^2}\sqrt{2R^2 - 2M} 
   = -\frac{1}{\kappa} \sqrt{-\tfrac{\Lambda}{6} (1-\tfrac{1}{12}M\Lambda )}
\ee
Note that in the limit $M \rightarrow 0$, we recover the Randall-Sundrum 
relation between $V_0$ and $\Lambda$ \cite{RS1, RS2}\footnote{Also the limit 
$R\rightarrow \infty$ is well defined, recovering the standard Schwarzschild 
solution in Minkowski space, where the relation between $r$ and $\bar{z}$ is: 
$\bar{z}=z_0-\sqrt{r^2-2M}$. }. 

In summary what we have done is basically to consider Schwarzschild-AdS 
space time with a brane located at a given distance $r(0)$ from the event
horizon. Then replace the part of the space time outside the brane ($r >
r(0)$) with a copy of the inner part, ending up with a finite range for
the radial variable. It is important to note that this space time comes
with two space-like singularities hidden inside the event horizons.  
Comparison with the dilatonic solution found on
previous work \cite{GJS}, shows that the role of the singularity on
those solutions is replaced by the event horizon in this new model.
Nevertheless we also have a non isotropic worldbrane, the time direction
scales differently than the space directions under radial
flow.
\begin{figure}
\begin{center}
\leavevmode
\epsfxsize=10cm
\epsffile{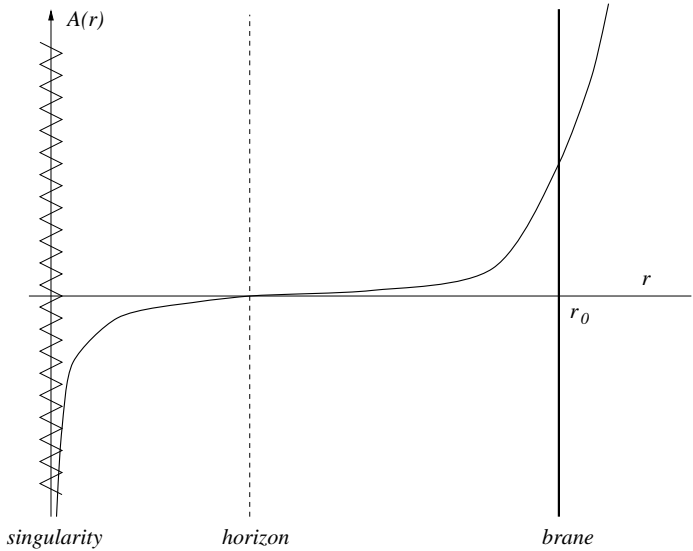}
\caption{\it Graviton profile on the Schwarzschild-AdS space-time. For large 
$r$ the graviton behaves like in ordinary AdS space and is not normalizable. A 
cutoff in form of a brane is needed at $r=r_0$. For small values of $r$, the 
graviton also diverges, but is hidden behind a horizon.}
\label{gravit}
\end{center}  
\end{figure}
\begin{figure}
\begin{center}
\leavevmode
\epsfxsize=10cm
\epsffile{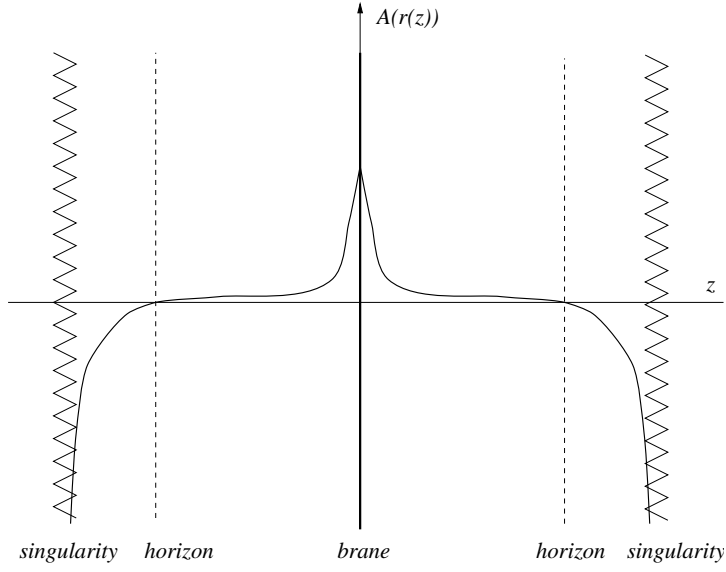}
\caption{\it The profile of normalized graviton after the cut off. The thrown 
away part is replaced by a copy of the space with $z<z_0$. The graviton is 
localized around the brane. The space-time ends in two singularities which do 
not harm the causal structure of the brane world, since they are hidden by  
event horizons. }
\label{metric}
\end{center}  
\end{figure}
\vspace{.5cm} 

\noindent
{\bf 3. Gravitational perturbations} \\
\vspace{.3cm}

To calculate the behavior of the graviton we add small fluctuations 
$h_{\mu\nu}$ to the above background, choosing the following gauge:
\be
\hg_{\mu\nu}= g_{\mu\nu} + h_{\alpha\beta} \delta^\alpha_\mu \delta^\beta_\nu, 
\ee
where $\alpha, \beta$ run over the coordinates $t$ and the angular coordinates
 $x^m$. Furthermore we have $h= g^{\mu\nu} h_{\mu\nu}= A^{-2} h_{tt} +
B^{-2} \bar{h}$ and $\nabla h=0$, $\bar{\nabla}^mh_{m\mu}=0$, where $\bar{\nabla}_m$ 
stands for the covariant derivative of the angular coordinates. Notice that 
this is not the usual de Donder gauge since the perturbation is not traceless.
Nevertheless if we are interested in a real graviton with two helicity states 
more constraints should be added. 
 
The equation of motion, on this gauge for the fluctuation  $h_{\mu\nu}$ is:
\be
\nabla^2 h_{\mu\nu} - 2 \nabla^\rho \nabla_{(\mu} h_{\nu)\rho} 
       + \tfrac{1}{3} \Lambda \ h_{\mu\nu} - \half \Lambda g_{\mu\nu} h= 0 \ ,
\ee

Introducing the background (\ref{Ansatz}), the components $\{tt\}$ and $\{zz\}$
of the above equation, reduce to:
\bea
&& A^{-3} A' \part_z h_{tt} - 2 A^{-4}(A')^2 h_{tt} + 
             B^{-3} B' \part_z \bar{h} - 2 B^{-4}(B')^2\bar{h} =0 \nn
&& \part^2_z h_{tt} + A^{-2} \part^2_t h_{tt} - B^{-2} \bar{\nabla}^2 h_{tt}
                 3(B^{-1}B' - A^{-1}A')\part_z h_{tt} \nn
&& \hspace{4.3cm}
            + (4A^{-2} (A')^2 - \tfrac{1}{3}\Lambda ) h_{tt} + \half A^2 h=0  
\eea
To describe four-dimensional zero modes, we consider
eigenfunction of the world brane variables $x^\alpha$, 
satisfying
\be
A^{-2} \part^2_t h_{tt} - B^{-2} \bar{\nabla}^2 h_{tt}= 0.
\ee
Under
these conditions we find a very simple solution:
\be
h_{tt} = A^2, 
\hspace{2cm}
\bar{h} = B^2
\ee
In principle we could turn on more degrees of freedom, to determine a more 
realistic graviton, nevertheless this mode shows the correct behavior to 
illustrate the location of the perturbation, and its normalizability.
Note that the specific form of our perturbation reproduces the desired
localization on the brane (see fig. 1,2) as well as the normalizability
condition.
              
To end this letter we would like to relate the world brane Newton constant
with the five dimensional Newton constant. To proceed on this direction we
note that a straightforward definition is not possible since the obvious
Kaluza-Klein reduction gives no terms in the effective action that could
be related to the Einstein term. This is a
consequence of the anisotropy of the world brane. Fortunately far from
the event horizon this space time looks like AdS, therefore our brane
becomes isotropic with warp factor $A^2$. Then 
we can proceed as usual to define the Newton constant.

The Newton constant, far from the horizon is essentially AdS in static
coordinates plus a correction coming from the black hole:
\be
M_{4}^2 = M_5^3\int_0^{z_0} dz \ A^2 (r(\bar{z}))
                = r_0 \ \Bigl(1 + \tfrac{r^2_0}{R^2} +
\tfrac{2M}{r^2_0}\Bigr)
\ee
For $r_0=R$ and $M\ll R$
\be
M_{4}^2 = M_5^3 \sqrt{\frac{-64}{3\Lambda}}\
                          \Bigl( 1- \tfrac{1}{8}M \Lambda \Bigr)
\ee

Again the warp factor defines the hierarchy and goes like $A^2$.`

\vspace{1cm}
\noindent
{\bf Acknowledgments}\\
We thank E. Alvarez for helpful conversations. The work of C.G. and B.J. has been supported by the TMR program FMRX-CT96-0012  on {\sl Integrability, 
non-perturbative effects, and symmetry in quantum field theory}. The work of P.S. was partially supported by the government of Venezuela.


\end{document}